\documentclass[pra, amsfonts, amssymb, amsmath,doi,reprint, showkeys, twoside,nobibnotes,floatfix]{revtex4-1}
\usepackage[colorinlistoftodos, color=green!40, prependcaption]{todonotes}
\usepackage{hyperref}

\usepackage{amsthm}
\usepackage{mathtools}
\usepackage{physics}
\usepackage{xcolor}
\usepackage{graphicx}
\usepackage{adjustbox}
\usepackage{placeins}
\usepackage[T1]{fontenc}
\usepackage{lipsum}
\usepackage{csquotes}

\usepackage{url}
\usepackage{graphicx}
\usepackage{amsmath}
\usepackage{bm}
\usepackage{braket}
\usepackage[]{algorithm2e}

\begin{document}
%
\title{Towards High Performance Relativistic Electronic Structure Modelling: \\ The EXP-T Program Package}
%
%
\author{Alexander V. Oleynichenko$^{1,2}$}
\email{alexvoleynichenko@gmail.com}
\homepage{http://www.qchem.pnpi.spb.ru}
\affiliation{$^{1}$Petersburg Nuclear Physics Institute named by B. P. Konstantinov of NRC ``Kurchatov Institute'', Gatchina, Russia \\ $^{2}$Department of Chemistry, Lomonosov Moscow State University, Moscow, Russia}
\author{Andr\'ei Zaitsevskii$^{1,2}$}
\affiliation{$^{1}$Petersburg Nuclear Physics Institute named by B. P. Konstantinov of NRC ``Kurchatov Institute'', Gatchina, Russia \\ $^{2}$Department of Chemistry, Lomonosov Moscow State University, Moscow, Russia}
\author{Ephraim Eliav$^{3}$}
\affiliation{$^{3}$School of Chemistry, Tel Aviv University, Tel Aviv, Israel}

\date{07.04.2020}

\begin{abstract}
Modern challenges arising in the fields of theoretical and experimental physics require new powerful tools for high-precision electronic structure modelling; one of the most perspective tools is the relativistic Fock space coupled cluster method (FS-RCC). Here we present a new extensible implementation of the FS-RCC method designed for modern parallel computers. The underlying theoretical model, algorithms and data structures are discussed. The performance and scaling features of the implementation are analyzed. The software developed allows to achieve a completely new level of accuracy for prediction of properties of atoms and molecules containing heavy and superheavy nuclei.
\end{abstract}


\maketitle              


%
\section{Introduction}

Nowadays first-principle based electronic structure modelling is widely recognized as a powerful tool for solving both fundamental and applied problems in physics and chemistry \cite{Dykstra2005}. A bulk of modern experiments in fundamental physics employing atomic and molecular systems seem to be hardly implementable or even senseless without theoretical predictions and assessments; some recent and the most striking examples are experiments for the electron electric dipole moment search \cite{Petrov2017}, design of laser-coolable molecular systems \cite{Ivanov2019} and spectroscopy of short-lived radioactive atoms and molecules \cite{Laatiaoui2016,GarciaRuiz2019}. Probably the most intriguing applications of quantum chemical modelling to fundamental problems are associated with molecules containing heavy and superheavy elements \cite{Eliav2015}; these applications require theoretical predictions to be accurate enough to be useful. For example, recent spectroscopic investigations of short-lived radioactive systems (No, Lr, RaF) required predicted excitation energies to be accurate up to 200-500 cm${}^{-1}$ in order to plan spectroscopic experiment, reduce its cost crucially and decode experimentally observed spectrum. Such an outstanding accuracy is unreachable without careful treatment of the so-called relativistic effects, completely changing even the qualitative picture of electronic states and properties \cite{Dyall2007}.

One of the most promising electronic structure models suitable for solution of such problems is the relativistic coupled cluster (RCC) theory \cite{Visscher1996} and its extensions to excited electronic states \cite{Eliav1998,Visscher2001}. Despite such advantages of these methods as correct physical behaviour, conceptual simplicity and controllable accuracy, rather severe drawbacks are to be mentioned. The most important ones are the restricted scope of applicability (not all types of electronic states are accessible at the moment) and high computational
cost, at least $N^6$ ($N$ is
a system size parameter), for compact systems where no advantages can be taken from localization techniques \cite{Saitow2017}.
The former obstacle seems to be surmountable at least for systems with three open shells (unpaired electrons); further theoretical developments are required to overcome limitations of currently used models.

A crucial step towards high precision relativistic modelling of molecular systems was made in the frames of the DIRAC project \cite{DIRAC}. Within this project, the wide variety of relativistic electronic structure models was developed and implemented as the modern and rather
efficient program package. However, the design of RCC codes implemented there seems to be not flexible enough to be able to construct the new
more extended
generation of coupled cluster models, e.g. models with inclusion of triple excitation and/or
more than two open shells.
Important requirements for the modern computer implementation of RCC-like models are (a) subroutines should be organized into well-tested elementary blocks which allow working with operators of arbitrary excitation rank; (b) algorithms should be highly scalable and parallelizable with the possible lowest time complexity.

In this paper we discuss the general strategy of building the high performance relativistic coupled cluster code
and report first benchmarks of the newly developed EXP-T program package implementing the considered concepts
and algorithms.

\section{General Considerations}

\subsection{Relativistic Fock space coupled cluster method}

The Fock space (FS) RCC computational scheme implies the conversion of the relativistic many-electron Hamiltonian into the second quantized form
\begin{equation}
H=\sum_{pq} h_{pq} \{ a^{\dag}_p \, a_q \} +\frac{1}{4} \sum_{pqrs} V_{pqrs} \{ a^{\dag} _p\, a^{\dag} _q\, a_s\, a_r \}
\label{ham}
\end{equation}
where $a^{\dag}_p$, $a_q$ denote creation/destruction operators associated with one-electron functions (molecular spinors) and curly braces mark normal ordering with respect to some closed-shell Fermi vacuum determinant; coefficients $ h_{pq}$ , $V_{pqrs}$ are molecular integrals in the basis of these spinors. Molecular spinors are normally generated by solving Hartree--Fock-like equations for the vacuum determinant. The conventional FS-RCC version \cite{Eliav1998,Visscher2001,Kaldor1991} is based on defining complete model spaces \emph{via} the choice of ``active'' (valence) spinors and constructing the normal-ordered exponential wave operator,
\begin{equation}
\Omega=\left\{\exp(T)\right\}, \quad\; T=\sum_{pq\dots rs\dots} \, t_{pq\dots rs\dots}
\{ a^{\dag} _p\, a^{\dag} _q\,\dots\, a_s\, a_r \}
\label{woper}\end{equation}
where  $t_{pq\dots rs\dots} $ are cluster amplitudes and the summation is normally restricted to single and double
excitation operators (RCCSD) or additionally triple excitations (RCCSDT).
The wave operator should reconstruct the target many-electron wavefunctions from their model-space projections. Electronic state
energies and model-space parts of the corresponding wavefunctions are obtained as eigenvalues and eigenvectors
of the effective Hamiltonian
$H^{\rm eff}=\left(\overline{H \,\Omega}\right)_{Cl},$
 where the subscript $_{Cl}$ marks the closed (model-space) part of an operator and the overbar denotes its connected part.

Cluster amplitudes should satisfy the equations
\begin{equation}
t_{pq\dots rs\dots} =\displaystyle\frac{1}{D_{pq\dots rs\dots}}\left(\overline{V\,\Omega}-\overline{\Omega \,\left(\overline{V \,\Omega}
\right)}_{Cl}\right)_{pq\dots rs\dots},\;
\label{amplitudes}
\end{equation}
\begin{equation}
V=H-H_0.
\end{equation}
The subscripts $_{pq\dots rs\dots}$ in the r.h.s. indicate that
the excitation $ \{ a^{\dag} _p\, a^{\dag} _q\,\dots \, a_s\, a_r \}$
is considered. $H_0$ is the Hartree--Fock operator for the Fermi vacuum state and
the energy denominators $D_{pq\dots rs\dots}$ are the negatives of the differences of
$H_0$ eigenvalues associated with the excitation.

It is convenient to partition the cluster operator $T$ according to the number of valence holes ($n_h$) and valence particles ($n_p$) to be
created/destroyed (e.g, related to ($n_h$,$n_p$) sectors of the Fock space):
\begin{equation}
T=\sum_{n_h n_p } \, T^{(n_h n_p)}
\label{sectors}\end{equation}
To describe the electronic states
in the ($N_h$,$N_p$) sector of the Fock space,
one needs to determine only $T^{(n_h n_p)}$ with $n_h\le N_h$ and $n_p\le N_p$. Therefore
the system of coupled equations  (\ref{amplitudes})  is splitted into subsystems, which can be solved consecutively.

The straightforward application of the complete-model-space FS-RCC method to molecular excited state calculations is severely restricted by unavoidable (at least for certain
ranges of nuclear configurations) numerical instabilities of the solutions of Eq.~(\ref{amplitudes}) caused by intruder states \cite{Evangelisti1987}. The presence of intruder states normally
manifests itself as the appearance of small or positive $D_{pq\dots rs\dots}$ values in Eq.~(\ref{amplitudes}).
In Ref.~\cite{Zaitsevskii2017} we modified the conventional FS-RCC equations (\ref{amplitudes})
\emph{via}
introduction of the special shifting of the ill-defined (nearly zero or positive) denominators.
This stratagem enables one to obtain stable solutions of amplitude equations in
problematic situations.
Strongly affecting only highly excited approximate eigenstates, it enables one to achieve an accurate description of low-lying excited states~\cite{Zaitsevskii2017,Kozlov2020}. Moreover,
results can fe further rectified by extrapolation to the zero-shift limit~\cite{Zaitsevskii2018}.

\subsection{Algorithm design}\label{sec:algorithms}

\begin{figure}[h!]
\center
\includegraphics[height=9cm]{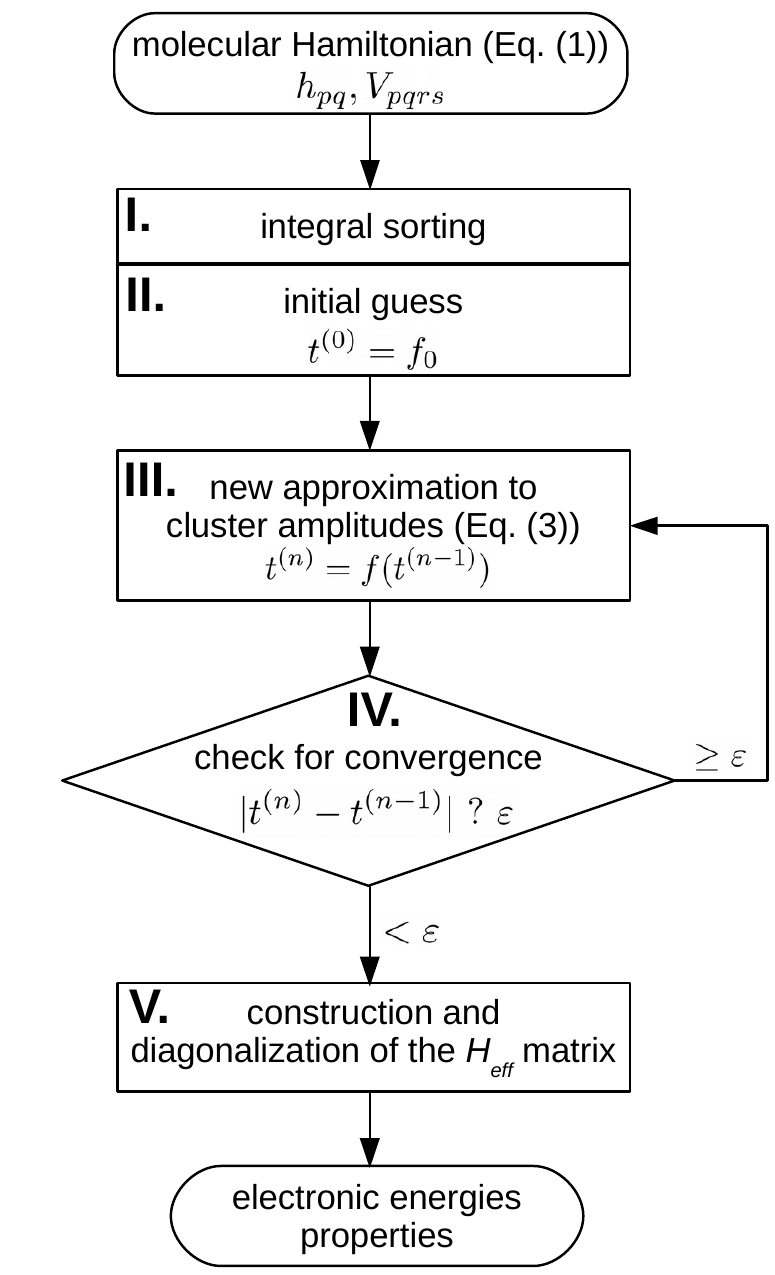}
\caption{Flowchart of the FS-RCC method. Steps I-V are performed consecutively for each Fock space sector.}
\label{fig:flowchart}
\end{figure}

The scheme of solving the working equations (\ref{amplitudes}) of the FS-RCC method can be formalized using the flowchart shown on Fig.~\ref{fig:flowchart}. Note that steps I-V are performed consecutively for each Fock space sector from the vacuum through the target one.

We proceed to the analysis of the time complexity of FS-RCC calculations. The usual
measure of the system's size in quantum chemistry is the number of one-electron basis functions $N$, roughly proportional to the number of atoms in the simulated system. Integral sorting (step I) requires $N^4$ operations; steps II -- IV require at least $N^6$ or even $N^8$ (for models involving triple excitations in Eq. (\ref{woper})) operations. Steps I, II and IV are
much cheaper than step III. Finally, the time complexity of step V is completely determined by the dimension of the subspace of active spinors and the number of active quasiparticles. There are numerous well-established and highly efficient algorithms for matrix diagonalization and the step V is the cheapest step of the FS-RCC calculation (at least for sectors with no more than three open shells). Further we will focus on the solution of amplitude equations, since this step is dominating at the RCCSD
and higher levels of theory.

Working equations of all RCC models are most conveniently formulated within the language of Goldstone diagrams \cite{Shavitt2009}. Cluster amplitudes can be represented as sums of dozens or even hundreds diagrams. It is worth noting that almost all diagrams for non-trivial FS sectors can be obtained from diagrams for the
conventional single-reference CC method simply by ``turning down'' open lines \cite{Kaldor1987}; this fact greatly simplifies validation of FS-RCC codes.

\begin{figure}[h!]
\center
\includegraphics[height=3cm]{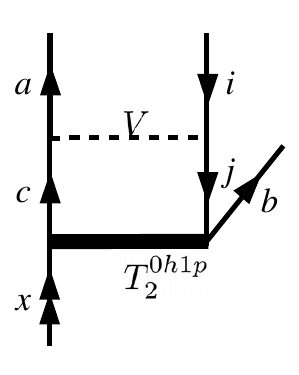}
\caption{Goldstone diagram for the contribution (\ref{eq:contraction}) to the $T_2^{(0h,1p)}$ operator amplitudes in the (0h,1p) sector.}
\label{fig:diagram}
\end{figure}

Fortunately, all the Goldstone diagrams in FS-RCC amplitude equations (\ref{amplitudes}) can be processed in a rather
similar way. Consider, for example, one of the simplest diagrams contributing to the $T_2^{(0h,1p)}$ cluster operator amplitudes (see Fig.~\ref{fig:diagram}). Its algebraic
expression is
\begin{equation}\label{eq:contraction}
- P(ab) \sum_{\bm{jc}} t_{x\bm{jc}b} V_{\bm{c}ia\bm{j}} \quad \text{for all} \ x,i,a,b
\end{equation}
where $t$ are cluster amplitudes and $P(ab)$ is a permutation operator. Here we use the widely accepted naming convention for the                                                                                         spinor indices \cite{Shavitt2009}: $i,j,...$ enumerates holes, $a,b,...$ -- particles, $x$ -- active particles. Indices to be contracted over are denoted with bold.

The straightforward summation is possible, but requires hand-coded loops and non-contiguous memory access, thus resulting in a confusing
algorithm and low performance.
It is preferable to split the evaluation of this expression into the sequence of elementary operations in such a manner that tensor contraction (\ref{eq:contraction}) can be performed as a simple matrix multiplication; in this case multidimensional arrays (which are commonly called tensors in many-body theories) are considered as rectangular supermatrices. These elementary operations will be:

\begin{enumerate}
\item $t_{x\bm{jc}b} \rightarrow t'_{xb\bm{jc}} \qquad\qquad\qquad\quad\ \ \forall\ x,j,c,b \qquad N^4$
\item $V_{\bm{c}ia\bm{j}} \rightarrow V'_{ia\bm{jc}}  \qquad\qquad\qquad\quad\ \ \forall\ c,i,a,j \qquad N^4$
\item $\sum_{\bm{jc}} t'_{xb;\bm{jc}} V'_{ia;\bm{jc}} \rightarrow \Delta t'_{xb;ia} \qquad \forall\ x,i,a,b \qquad N^6$
\item $\Delta t'_{xbia} \rightarrow \Delta t_{xiab} \qquad\qquad\qquad\ \forall\ x,i,a,b \qquad N^4$
\end{enumerate}

The most time consuming operation 3 (tensor contraction) can be performed as a matrix multiplication thus allowing to use any high-performance linear algebra package. Additional tensor transpositions (1, 2, 4) are now required
 for most diagrams. Such tensor transpositions in general case can hardly be implemented in a cache-efficient manner (except of purely 2D matrix transposition-like cases). However, for actual problems these three additional transpositions are necessarily cheaper than the tensor contraction step. This approach is sometimes referred as the Transpose-Transpose-\texttt{GEMM}-Transpose (TTGT) approach \cite{Matthews2018}. Furthermore, the two other important advantages of such a decomposition into elementary operations are to be mentioned here:

(1) only these operations are to be implemented for arbitrary rank tensors; the code for all CC models can be in principle obtained in an automated manner. This ensures flexibility and extensibility of the code written in this elementary building blocks paradigm;

(2) these elementary operations are perfectly suitable for parallel execution.

\subsection{Symmetry handling and data structures}\label{seq:datastruct}

Below a brief discussion of data structures optimally compatible with the algorithms described above and ensuring efficient and well-scaling parallel implementation on heterogeneous architectures is presented (in fact, all modern supercomputers are of this type).
The basic idea is to choose some partitioning of all the data (e.g. cluster amplitudes and molecular integrals) to be processed into blocks. The most computationally feasible way of such a partitioning is determined by division of the whole range of molecular spinors into subsets; the resulting tensors can be considered as generalizations of block matrices. In case of additional spatial symmetry, it is natural to place the spinors which transform via the same irreducible representation into the same subset, thus allowing to get rid of matrix elements which are \textit{a priori} zero due to symmetry reasons. This approach is known as the direct product decomposition (DPD) technique and is widely used in both non-relativistic \cite{Stanton1991} and relativistic \cite{Shee2016} frameworks. Note that in the general case complex arithmetic is required in relativistic electronic structure calculations; however, for some double groups (e.g. $C_{2v}^{*}$, $C_{2h}^{*}$, $D_{2}^{*}$, $D_{2h}^{*}$, $C_{\infty v}^{*}$, $D_{\infty h}^{*}$) real arithmetic can be used \cite{Saue1999}, resulting in a great reduction of computational effort and memory requirements.

\begin{figure}[h!]
\center
\includegraphics[width=\columnwidth]{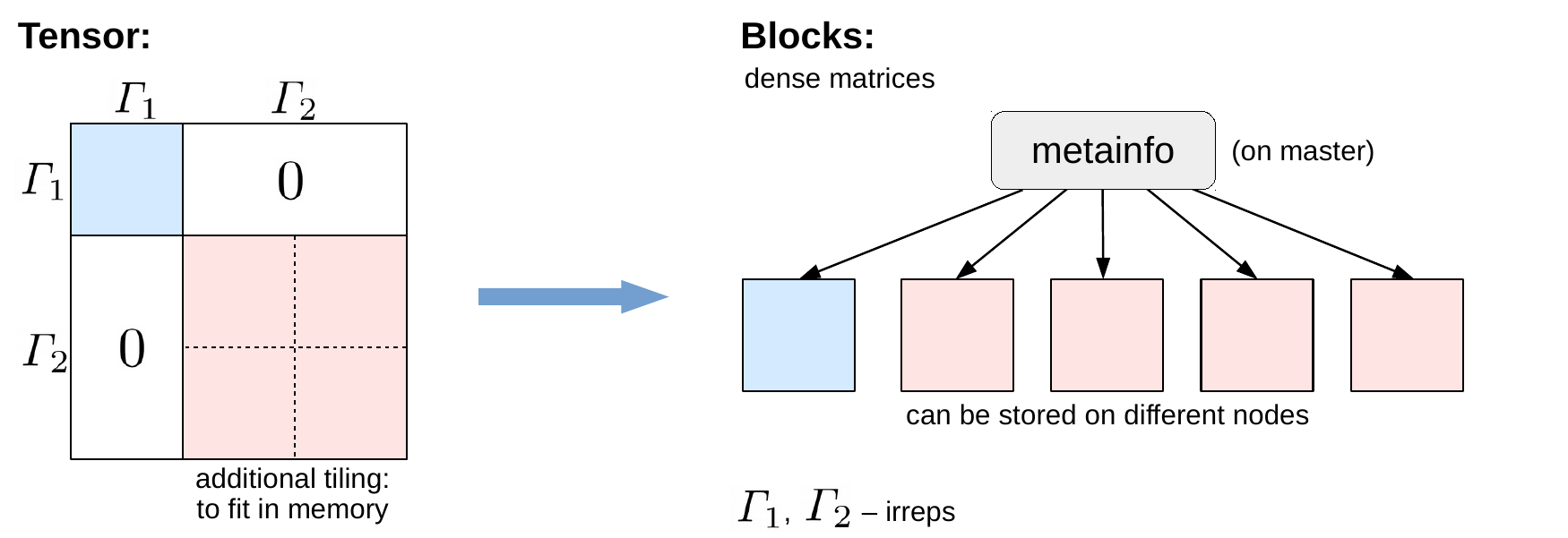}
\caption{Partitioning of tensors (cluster amplitudes or molecular integrals) into smaller blocks. The figure shows the case of a rank 2 tensor; generalization to the case of rank 4 and 6 tensors is straightforward.}
\label{fig:blocks}
\end{figure}

\begin{figure}[h!]
\center
\begin{algorithm}[H]
 \KwData{tensors $A$, $B$}
 \KwResult{tensor \\ $C = \sum_{\bm{k_1,...,k_K}} A_{i_1,...,i_M;\bm{k_1,...,k_K}}$, $B_{j_1,...,j_N;\bm{k_1,...,k_K}}$}
 \ForEach{block\_c in blocks(C)} {
 \ForEach{block\_a in blocks(A)} {
 load block\_a from disk if needed\;
 \ForEach{block\_b in blocks(B)} {
    load block\_b from disk if needed\;
    calculate dimensions of supermatrices\;
 	block\_c += \texttt{gemm}(block\_a, block\_b)
 }
 }
 store block\_c to disk if needed\;
 }
\end{algorithm}
\caption{Contraction of two tensors stored as lists of dense blocks. Loops can be swapped to ensure that the outer loop runs over the tensor stored on disk (to avoid redundant disk operations).}
\label{fig:mult}
\end{figure}

Thus all tensors can be represented as lists of blocks (Fig.~\ref{fig:blocks}); all algorithms of elementary operations described in~\ref{sec:algorithms} are expressed in terms of these blocks (for example, see Fig.~\ref{fig:mult} for the tensor contraction algorithm). Blocks of one tensor are independent and can be stored on different nodes of the distributed memory system thus reducing memory requirements and allowing treatment of really large systems.

The node-level parallelism (OpenMP\cite{Dagum1998} or GPGPU) arises naturally for elementary operations with blocks. Current computer implementation of FS-RCC reported here uses highly optimized MKL \cite{MKL2018} and CUBLAS \cite{CUBLAS} libraries to perform parallel contractions on CPU and GPU, respectively. It should be noted that sizes of blocks for high symmetry point groups ($C_{\infty v}^{*}$ represented by $C_{32}$ and $D_{\infty h}^{*}$ represented by $C_{16h}$) can differ by orders of magnitude. This gives rise to a considerable imbalance:
strong scaling (wrt number of threads) will be
efficient for large blocks and
inefficient for relatively small ones, leading to the degradation of overall scalability of the program.
This issue is not addressed here, but the
obvious solution can be based on the dynamic selection of optimal number of threads guided by the runtime profiling performed at the first iteration.

\section{Implementation and Benchmark}

\subsection{The EXP-T program system}

Considerations discussed in Sect.~\ref{sec:algorithms} and~\ref{seq:datastruct} were implemented in the newly developed electronic structure package EXP-T (named after the formula (\ref{woper}) of the RCC Ansatz).

Parameters of the relativistic Hamiltonian (\ref{ham}), i.e. sets of molecular spinors and molecular integrals, have to be imported from third party electronic structure packages. Currently EXP-T is interfaced to the DIRAC program package \cite{DIRAC}, thus getting access to the wide variety of Hamiltonians (e.g. four-component ones and relativistic pseudopotentials) and property operators implemented there. All RCC codes are Kramers unrestricted.

Electronic structure models available in EXP-T are listed below:
\begin{itemize}
\item single-reference CCSD, CCSD(T), CCSDT-n (n=1,2,3), CCSDT models;
\item FS-CCSD model for the (0h,1p), (1h,0p), (1h,1p), (0h,2p), (2h,0p), (0h,3p) Fock space sectors;
\item FS-CCSDT-n (n=1,2,3) and FS-CCSDT models are implemented for the (0h,1p), (0h,2p) and (0h,3p) Fock space sectors.
\end{itemize}
At present only single-point energy calculations are implemented for all the models listed above. The FS-RCC models for the (0h,3p) FS sector were developed to deal with electronic states dominated by determinants with three open shells and was implemented earlier only by Kaldor and coworkers with application to very small non-relativistic atomic systems \cite{Hughes1995}; the features of these models
will be described in our future papers. The corresponding code is to be considered as the experimental to the moment.

Some features recently proposed by us which greatly extend the scope of applicability of the FS-RCC method are also included:

\begin{itemize}
\item ``dynamic'' energy denominators shifts and subsequent Pad\'e extrapolation to the zero-shift limit as a solution of the intruder-state problem \cite{Zaitsevskii2017,Zaitsevskii2018};
\item finite-field transition property calculations \cite{Zaitsevskii2018a};
\item decoupling of spin-orbit-coupled states by projection and extraction of SO coupling matrix elements \cite{Zaitsevskii2017}.
\end{itemize}

EXP-T currently supports parallel calculations on shared-memory computers via the OpenMP and CUDA technologies.

EXP-T is written in the C99 programming language and hence can be compiled using the most common development tools available on most platforms. EXP-T is currently oriented to Unix-like operating systems. 

\subsection{Performance evaluation}

To assess the performance features of the newly developed FS-RCC implementation, the series of the FS-RCCSD calculations of the KCs alkali-metal molecular dimer were done.
In order to test the efficiency of the blocking scheme employed we performed the calculations with full account for the point-group symmetry and with artificially lowered symmetries. The size of the problem (112 active spinors, 374 spinors overall) is large enough to demonstrate tendencies in scaling features and relative computational cost of different stages of the routine FS-RCC calculation. Electronic states of KCs
are formed by the following Fock space scheme:

\[ \rm KCs^{2+} (0h,0p) \rightarrow \rm KCs^{+} (0h,1p) \rightarrow \rm KCs^{0} (0h,2p) \]

Results of wall time measurements for the RCCSD(0h,0p) calculation of KCs${}^{2+}$ in different point groups are presented in Tab.~\ref{tab:time}. The speed-up is rising from $C_1^{*}$ to $C_s^{*}$ and from $C^{*}_{2v}$ to $C^{*}_{\infty v}$ (represented actually by $C_{32}$) and comes from the reduction of tensor sizes; the speed-up going from $C_{s}^{*}$ to $C_{2v}^{*}$ is only due to the use of real arithmetic instead of complex one since both groups have the same number of fermionic irreducible representations (two) (see also \cite{Visscher1996a}). It should be mentioned that the $C_1^{*}$ point group represents the most general case and probably 
will be the most demanded in future applications of the FS-RCC method to polyatomic molecules.

\begin{table*}\label{tab:time}
\caption{Wall clock time (in seconds on the Intel(R) Xeon(R) E5-2680 v4 CPU) for RCCSD calculation in different point groups. The test calculation concerns the KCs${}^{2+}$ ion in the molecular pseudospinor basis (16 electrons, 374 functions). $R$ -- real group, $C$ -- complex group.}
\begin{center}
\begin{tabular*}{\textwidth}{lc @{\extracolsep{\fill}} rrrrr}
\hline
Point & & Total & Integral & Tensor & Tensor & Time per \\
group & & time  & sorting  & contractions & transpositions & iteration \\
\hline
$C_1^*$			&$(C)$&	105694	&	19913	&	37375	&	2122	&	2218		\\
$C_s^*$			&$(C)$&	25135	&	11198	&	12507	&	1049	&	786			\\
$C_{2v}^*$		&$(R)$&	11489	&	5525	&	5307	&	471		&	324			\\
$C_{\infty v}^*$&$(R)$&	4113	&	2442	&	1451	&	137		&	87			\\
\hline
\end{tabular*}
\end{center}
\end{table*}

Furthermore, relative computational costs of stages of the FS-RCC calculation are nearly constant for different point groups. Integral sorting being the $N^4$ operation requires considerable computational time, but its contribution will decrease for larger problems.
Moreover, the tensor contractions / tensor transpositions ratio is high enough to completely justify the use of the decomposition (TTGT) approach to evaluation of the general tensor contractions presented in Sect.~\ref{sec:algorithms}. This ratio will be much higher for larger  problems due to the ratio of time 
scaling of these tensor operations ($N^{4-6}$ \textit{vs} $N^{6-8}$).

\begin{figure*}
\center
\includegraphics[width=0.85\textwidth]{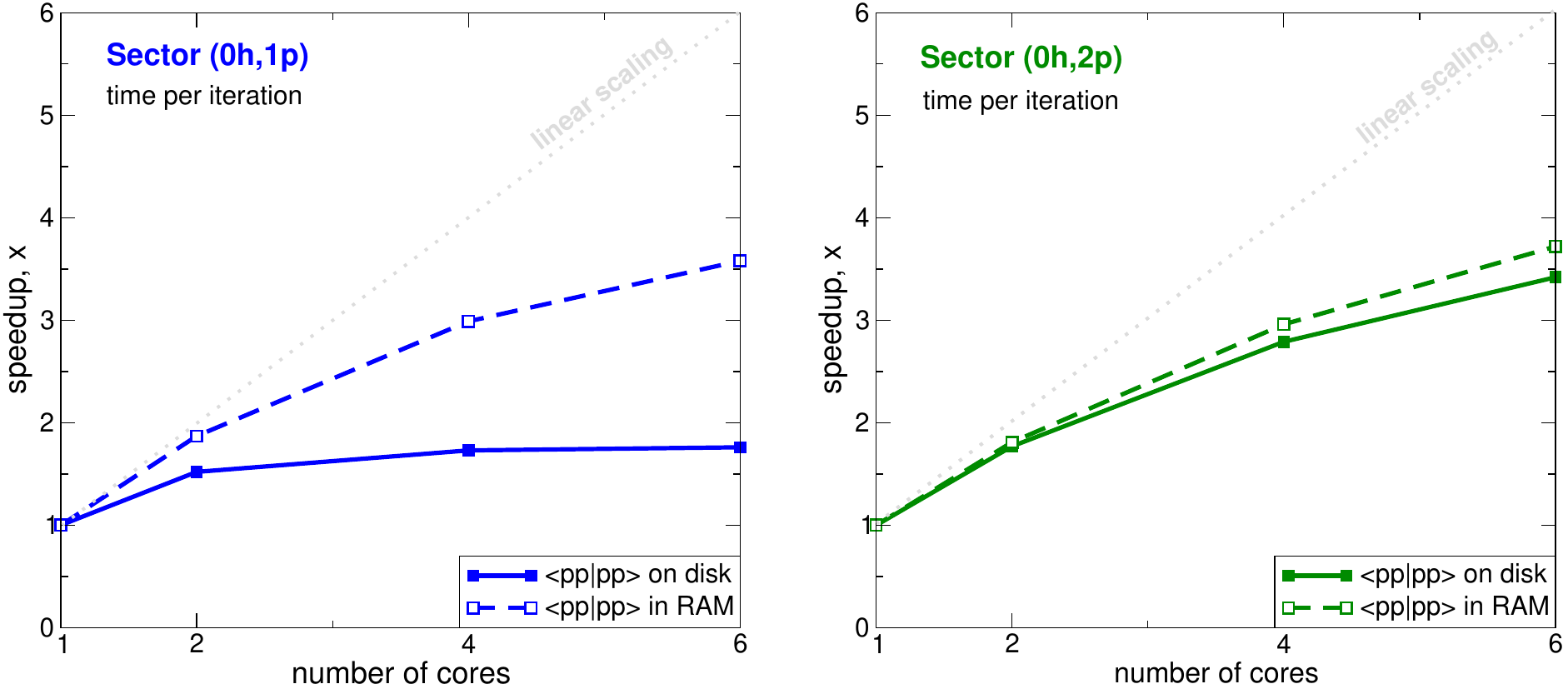}
\caption{Scaling of the FS-CCSD calculation in different Fock space sectors with respect to the number of OpenMP threads (for the time per iteration on the Intel(R) Core(TM) i7-4930K CPU). The test calculation concerns the KCs molecule in the molecular pseudospinor basis (16 electrons, 374 functions, 112 active spinors).}
\label{fig:openmp}
\end{figure*}

Nearly 90\% of time used for tensor contractions is spent for evaluation of the 
single term in RCC equations which involves two-electron integrals with four indices of virtual spinors. This tensor (commonly denoted as $\braket{pp|pp}$) is typically an order of magnitude larger than the other tensors containing integrals and has to be stored on harddrive in most real-life applications; this tensor is to be read from disk at each CC iteration (see Fig. \ref{fig:mult}), resulting in limited parallelizabitity of the code (Fig. \ref{fig:openmp}). The possible remedy is to store some part of the $\braket{pp|pp}$ tensor in RAM.

The better scaling for the (0h,2p) Fock space sector than for the (0h,1p) sector is due to the amount of data processed in the former case is much 
bigger: in the example considered the maximum block size is of order $1 \cdot 10^6$ elements for the (0h,1p) sector and $5 \cdot 10^6$ for the (0h,2p) sector. Thus the percentage of computations that can be performed in parallel is considerably higher for the latter case resulting in better scaling (due to the Amdahl's law). Another problem which can lead to worse scaling is the fact that for highly symmetrical point groups like $C_{\infty v}^{*}$ sizes of blocks can be very small. This results in overheads for thread creating being much larger than the time used for actual calculation and hence in degradation of the overall scaling. Such a situation is observed for the (0h,2p) sector (Fig.~\ref{fig:openmp}, right plot). The latter problem can be solved by 
choosing optimal number of threads for such small blocks.

\section{Conclusions and Prospects}

A new implementation of the Fock space relativistic coupled cluster method designed for modern parallel systems is presented; underlying method, algorithms and approaches to data handling are discussed. Scaling with respect to the number of OpenMP threads 
currently is not ideal (in the example presented no more than 4x times faster on 6 CPU cores), but the ways of possible improvements are rather clear. However, conceptual limitations due to the necessity of the usage of harddrive to store molecular integrals can be overcame in future versions of the code by employing the MPI parallelization model.

The future work on the EXP-T program system will address not only improvements of the computational scheme, but also development of new relativistic coupled cluster models aimed at expanding the field of applicability of the FS-RCC method and achieving a 
principally new level of accuracy for prediction of properties of molecules containing heavy and superheavy nuclei.

\section{Acknowledgements}

Authors are grateful to T. A. Isaev, S. V. Kozlov, L. V. Skripnikov, A. V. Stolyarov and L. Visscher for fruitful discussions. This work has been carried out using computing resources of the federal collective usage centre Complex for Simulation and Data Processing for Mega-science Facilities at NRC ``Kurchatov Institute'', \url{http://ckp.nrcki.ru/}, and computers of Quantum Chemistry Lab at NRC ``Kurchatov Institute'' -- PNPI.

The research was supported by the Russian Science Foundation (Grant No. 20-13-00225).


\bibliography{aoleynichenko}

\end{document}